\documentclass[prd,amsmath,amssymb,]{revtex4}
\usepackage{graphicx}
\usepackage{color}
\usepackage[colorlinks=true,linkcolor=red]{hyperref}
\usepackage{hyperref}

\begin{document}
\title{The equivalence principle, uniformly accelerated reference frames, and the uniform gravitational field}
\author{Gerardo Mu{\~n}oz}
\email{gerardom@csufresno.edu}
\affiliation{Department of Physics, California State University, Fresno, Fresno, CA 93740-8031}
\author{Preston Jones}
\email{prjones@ulm.edu}
\affiliation{Department of Mathematics and Physics, University of Louisiana, Monroe, LA 71209-0575}

\date{\today}

\begin{abstract}
The relationship between uniformly accelerated reference frames in flat spacetime and the uniform gravitational field is examined in a relativistic context. It is shown that, contrary to previous statements in the pages of this journal, equivalence does not break down in this context. No restrictions to Newtonian approximations or small enclosures are necessary.
\end{abstract}

\maketitle

\section{Introduction}

It would not be very controversial to state that there is universal agreement on the meaning and implementation of the equivalence principle within a Newtonian framework. As a review of the extensive literature on the subject will easily confirm, however, the situation changes drastically in the relativistic domain. Most authors correctly point out that, in general, the equivalence principle only applies locally,\cite{Moller, Schutz, MTW, Stephani, Rindler, Carroll} but there are several others who deny its validity even at the local level.\cite{Synge, Fock, Ohanian} The disagreement shows up not only in discussions of the foundations of general relativity, but also in applications such as the radiation from accelerated charges\cite{Pauli,FultonRohrlich}, Unruh radiation\cite{Narozhny, FullingUnruh,Crispino}, and uniform gravitational fields\cite{Rohrlich, Desloge, Gron}, to name just a few. Our aim in this paper is to present a compelling demonstration that a relativistic uniformly accelerated reference frame in flat spacetime is equivalent to a uniform gravitational field, as one would naively expect from the nonrelativistic version of the equivalence principle. It might be argued that this issue was settled by Rohrlich\cite{Rohrlich} in a definitive manner a long time ago, but a paper by Desloge\cite{Desloge} in this journal revisited the issue with the conclusion that ``in a uniform gravitational field the space-time is as expected curved." Because of its obvious importance, we believe that a careful reexamination of the problem is necessary.

Most of our discussion will be easier to follow if the reader keeps a few simple facts about relativistic metric theories of gravity in mind.

F1. Imposing a-priori meanings on the coordinates in a metric theory of gravity is to be done at one's own peril. It is not uncommon to find that our preconceived notions fail in certain cases, or that our demands overconstrain the system.

F2. Acceleration -- either in the nonrelativistic form $d^2 x /dt^2$ or in the special-relativistic form $d^2 x^\mu /d \tau^2$ -- does not have an absolute (geometric) meaning in a metric theory of gravity. In other words, unlike the 4-velocity $d x^\mu /d \tau$, the object $d^2 x^\mu /d \tau^2$ is not a 4-vector under general coordinate transformations, but rather an intrinsically observer-dependent notion.

F3. One should never expect the relativistic description of a system to satisfy nonrelativistic constraints.

These facts have been phrased in terms of a metric theory of gravity rather than general relativity in order to emphasize their independence from any particular set of dynamical equations. As the reader will notice, Sections II and III below hold in any metric theory of gravity; only when we get to Section IV will we require Einstein's equations.

\section{Intuitive attempts at defining a uniform gravitational field}

In Ref.\cite{Desloge}, the metric ($c=1$ hereafter)
\begin{equation}
ds^2 = \alpha^2 (x) \, dt^2 - dx^2
\end{equation}
was motivated by the desire to represent a rigid frame in a gravitational field and a rigid frame uniformly accelerating in field-free space (the restriction to two-dimensions is of no consequence at this point; we shall take up the four-dimensional problem in later sections). It is clear that, as intended, the $x$-coordinate has been given an a-priori meaning such that the length of an infinitesimal object equals its coordinate length $dx$, i.e., lengths measured with this coordinate system are given by their special-relativistic inertial values. Desloge defines the ``local velocity" $\alpha^{-1} dx / dt$ and ``local acceleration" $\alpha^{-1} d \, (\alpha^{-1} dx / dt)/dt$ and uses the latter together with the equation of motion of a free particle to identify $\alpha^{-1} d \alpha / dx$ as the proper acceleration of an observer at rest at $x$ in the gravitational field. Taking the requirement that this acceleration be constant everywhere as a definition of the uniform gravitational field then leads to the conclusion that the Riemann tensor has nonvanishing components -- hence his statement that ``in a uniform gravitational field the space-time is as expected curved."

While Desloge's ``local acceleration" is a possible definition of acceleration, it is not mandated by general relativity or even special relativity. Moreover, it is an ill-motivated definition, since $\alpha \,dt$ is the time interval between two events occurring at a single spatial location ($dx=0$), and hence bears no special significance for the motion of a freely-falling particle because in that case we are interested in events that necessarily refer to different spatial locations. The point is best illustrated by recalling that already in special relativity one defines the acceleration by resorting to the particle's proper time rather than the proper time of the observer. The a-priori meaning of the spatial coordinate $x$ (recall fact F1) together with the arbitrary assignment of a preferred status to the ``local acceleration" (recall fact F2) collude to yield the erroneous conclusion that uniform gravitational fields must correspond to curved spacetimes. A few examples will help elucidate our claim. The geodesic equation of motion for an inertial observer is
\begin{equation}
{d^2 x^\mu \over d \tau^2} + \Gamma^\mu_{\nu \rho} \, {d x^\nu \over d \tau} \, {d x^\rho \over d \tau} =0
\end{equation}
With the metric (1) we find the two equations
\begin{eqnarray}
{d^2 t \over d \tau^2} + 2 \, {\alpha' \over \alpha} \, {d t \over d \tau} \, {d x \over d \tau} \!\! &=& \!\! 0 \\
{d^2 x \over d \tau^2} + \alpha \alpha' \left( {d t \over d \tau} \right)^2 \!\! &=& \!\! 0
\end{eqnarray}
Integrating equation (3) gives
\begin{equation}
{d t \over d \tau} = {k \over \alpha^2}
\end{equation}
with $k>0$ an integration constant which may be fixed by imposing initial conditions. Replacing this expression in Eq. (4) yields
\begin{equation}
{d \over d t} \left( {1 \over \alpha^2} \, {d x \over dt} \right) +  {\alpha' \over \alpha} =0
\end{equation}
which is equivalent to Desloge's Eq. (6).

The question before us now is how to {\sl define} a uniform gravitational field based on the acceleration of a test particle. One possibility is to argue that, since Newtonian physics must be exact for vanishing velocities and weak fields, we should concentrate on particles dropped from rest at time $t=0$ from a point $x=0$ where $\alpha (0 ) =1$ (this normalization is possible for an arbitrary point because the field is uniform) and define the acceleration in a small neighborhood of that event as $d^2 x /dt^2$. Equation (6) then implies
\begin{equation}
{d^2 x \over d t^2} = - \alpha \alpha' 
\end{equation}
in this small neighborhood. If we set the acceleration equal to the constant $-g$ we get the result
\begin{equation}
\alpha^2 = 1 + 2 gx
\end{equation}
As is well known, this is the correct nonrelativistic limit, but it should be clear from its derivation that this form of the metric cannot be expected to describe the field globally. Nevertheless, within its domain of validity, it represents a uniform gravitational field (in the intuitive sense that the acceleration equals $-g$), and corresponds to a flat spacetime, $R_{\mu \nu \kappa \rho} =0$, at $x=0$.

A second possibility is to define the relevant acceleration as $d \, (\alpha^{-1} dx / dt)/dt$. It then follows from (6) that
\begin{equation}
{d \over d t} \left( {1 \over \alpha} \, {d x \over dt} \right) +  \alpha'  =0
\end{equation}
at the initial instant, provided again that the particle is released from rest. One would therefore naturally choose to set $\alpha' =g$ and find as a consequence the Kottler-M\o ller\cite{Kottler1, Moller} metric
\begin{equation}
ds^2 = (1+gx)^2 \, dt^2 - dx^2
\end{equation}
The Riemann tensor associated with this metric vanishes everywhere.

These two simple examples are perhaps sufficient to prove the point that defining the uniform gravitational field based on the naive ``acceleration $=-g$" condition is not satisfactory. One of the main obstacles to this approach is clearly fact F2, and most of the rest of this paper will be dedicated to finding an unambiguous way of addressing this problem.

Before we proceed in that direction however, let us briefly explain the technical reasons why Desloge's ``local acceleration" fails -- as will almost every other definition -- to give a flat spacetime if we start with the metric (1). As Rohrlich\cite{Rohrlich} and Tilbrook\cite{Tilbrook} have shown using different methods, a metric of the form
\begin{equation}
ds^2 = \lambda^2(x) \, dt^2 - \sigma^2(x) \, dx^2
\end{equation}
is equivalent to a flat spacetime only if
\begin{equation}
\sigma(x) = {1 \over g} \, {d \lambda(x) \over dx }
\end{equation}
with $g$ an integration constant. Adopting a metric of the form (1) means choosing $\sigma =1$, which immediately turns $\lambda(x)$ into $\alpha(x) = 1+gx$, and the metric (11) into the Kottler-M\o ller metric (10). In other words, once we adopt coordinates such that the metric takes the form (1), we can either impose a condition on a given definition of acceleration (in which case we would find a spacetime which is, in general, not flat), or we can impose the condition that the spacetime should be flat (and find that the only definition of initial acceleration that works is $d \, (\alpha^{-1} dx / dt)/dt$). Imposing both conditions simultaneously will, in general, lead to incompatible demands on the single degree of freedom $\alpha(x)$ -- fact F1 once again. Thus, given the arbitrariness in the choice of coordinates and in the definition of acceleration, claiming that a uniform gravitational field  necessarily entails a curved spacetime is hardly a convincing argument if the uniform gravitational field is interpreted in the above sense of ``acceleration $=-g$" and we insist on imposing that condition on the metric (1). We emphasize again that these arguments are insensitive to the dynamics: the equations of general relativity have not been used anywhere so far.

A concrete example will probably illustrate the above conclusion best. One may ask whether it is possible to retain the concept of ``local acceleration" and still find a flat spacetime for a uniform gravitational field. To show that the answer is yes, write down the geodesic equation (2) for the metric (11). One finds that the equivalent to (4) may be put in the form
\begin{equation}
{1 \over \lambda} \, {d \over d t} \left( {1 \over \lambda} \, {d x \over dt} \right) + \left( {\lambda'' \over \lambda' } + G(x) \right)  \left( {1 \over \lambda} \, {d x \over dt} \right)^2 +  {g^2 \over \lambda \lambda' } = 0
\end{equation}
where
\begin{equation}
G(x) = - {1 \over \lambda} \, {d \lambda \over dx}
\end{equation}
Following Desloge we should interpret $-g^2 / \lambda \lambda' $ as the negative of the acceleration of the observer and consider
\begin{equation}
{g^2 \over \lambda \lambda' } = g
\end{equation}
as the definition of a uniform field. This condition is easily integrated to give $\lambda^2 = 1+2gx$, which, together with (12), leads to
\begin{equation}
ds^2 = (1+2gx) \, dt^2 - { dx^2 \over 1+2gx }
\end{equation}
This is the Kottler-Whittaker\cite{Kottler2, Whittaker} metric. It describes a flat spacetime and, in the present context, a uniform gravitational field since the ``local acceleration" is constant. A possible objection to our example may be that this metric does not correspond to a rigid frame since it does not have the form (1)\cite{Adler}. We shall dispense with this objection in the next section.

Finally, we mention for completeness a more sophisticated approach proposed by Rohrlich\cite{Rohrlich}. Since in special relativity the motion of a point particle with proper acceleration $-g$ obeys
\begin{equation}
{d \over dt } \left( \gamma \, {dx \over dt } \right) = -g
\end{equation}
with $\gamma = [ 1- (dx /dt)^2 ]^{-1/2}$ the usual gamma factor, we may ask whether the flat spacetime metric (11)-(12) is compatible with this condition. The geodesic equation (2) yields as before
\begin{equation}
{d t \over d \tau} = {k \over \lambda ^2}
\end{equation}
Choosing $\lambda(x=0) =1$ and an initial velocity $dx/dt =0$ at that point determines that $k=1$. Therefore $d \tau = \lambda^2 dt$, and (11) may be written as
\begin{equation}
\lambda^4 \, dt^2 = \lambda^2 \, dt^2 - \left( {1 \over g} \, {d \lambda \over dx }
\right)^2 dx^2 =  \lambda^2 \, dt^2 - {1 \over g^2} \, d \lambda^2
\end{equation}
or
\begin{equation}
d \lambda^2 = g^2 \lambda^2 (1-  \lambda^2 ) \, dt^2
\end{equation}
Furthermore, since the solution to (17) with the given initial conditions is
\begin{equation}
x = { 1 \over g }  \left( 1 - \sqrt {1+ g^2 t^2  } \right)
\end{equation}
we have
\begin{equation}
 {dx \over dt } = -{gt \over  \sqrt{ 1+ g^2 t^2  } } = -{\sqrt{ (1-gx)^2 -1 } \over 1-gx}
\end{equation}
so that, substituting for $dt$ in (20),
\begin{equation}
d \lambda = g \lambda \sqrt{ 1-  \lambda^2 } \; {1-gx \over \sqrt{ (1-gx)^2 -1 }} \, dx
\end{equation}
Integrating we find
\begin{equation}
\lambda (x) =  {1 \over \cosh \sqrt{ (1-gx)^2 -1 }}
\end{equation}
Computing $\sigma (x)$ from (12) yields the Rohrlich metric\cite{Rohrlich}
\begin{equation}
ds^2 =  {1 \over \cosh^2 \sqrt{ (1-gx)^2 -1 }} \left( dt^2 - {(1-gx)^2 \tanh^2  \sqrt{ (1-gx)^2 -1 } \over (1-gx)^2 -1 } \, dx^2 \right)
\end{equation}
This fairly complicated result seems to have the advantage of satisfying the intuitively reasonable condition (17). However, fact F2 and the non-inertial nature of the coordinates $x$ and $t$ should convince the reader that this apparent advantage is illusory.

\section{Accelerated frames in special relativity}

In this section we set aside the gravitational aspect of our problem and concentrate on the description of accelerated frames in flat spacetimes. As we shall see, a satisfactory resolution of the ambiguities that plagued the intuitive approach in the previous section cannot be reached without a deeper understanding of the role of acceleration in special relativity.

As stated previously, the trajectory of a point particle subject to an acceleration $g$ as measured in its instantaneous rest frame is determined by the equation of motion
\begin{equation}
{d \over dT } \left( \gamma \, {dX \over dT } \right) = g
\end{equation}
where $X$ and $T$ are the coordinates used by an inertial observer $O$, and the sign of the proper acceleration has been changed to suit the purposes of the present section. Arranging the initial conditions so that the particle is at $X_0$ with zero velocity at $T=0$ we find
\begin{equation}
X = { 1 \over g }  \left( \sqrt {1+ g^2 T^2  } -1 \right) + X_0
\end{equation}
Note that there is no ambiguity whatsoever in the interpretation of any quantity involved in (26) or (27) because the coordinates $X$, $T$ refer to an inertial frame.

A rigid frame is defined as a system in which all points remain at the same proper distance for all times. It follows immediately from (27) that these points cannot be moving with the same acceleration. Indeed, since the relative velocity vanishes at $T=0$, both the inertial and accelerated observers will, at that instant, measure the same spatial separation $\delta X_0$ between neighboring points in the accelerated frame (i.e., $\delta X_0$ is the proper distance between these points at $T=0$). As the velocity of the moving frame increases, Lorentz contraction sets in, and the proper length will be given by $\gamma \, \delta X$. But if all point move with the same acceleration $g$, (27) tells us that $\delta X = \delta X_0$ for all $T$, and the proper length between points in the accelerating frame would increase without bound as $T \rightarrow \infty$.

In order to find out how the acceleration should vary with the position $x$ in the accelerating frame it is helpful to introduce some conventions. We shall reserve the symbol $g$ for the proper acceleration of the origin of the accelerating frame, and use $a(x)$ for the proper acceleration of a point located at $x$ in that frame. If we let the origins of the two systems coincide at $T=0$, then (27) holds for $X_0 =0$, and
\begin{equation}
X = { 1 \over a }  \left( \sqrt {1+ a^2 T^2  } -1 \right) + X_0
\end{equation}
for the point $x$ such that $X_0= x$ at $T=0$.

If the accelerating frame is to move rigidly, we must satisfy the Lorentz contraction condition
\begin{equation}
\delta X(T) = { \delta x  \over \gamma (x) } 
\end{equation}
with the proper length $\delta x$ fixed for all times. Computing $\delta X(T)$ from (28) and $\gamma (x)$ from the derivative $dX/dT$ of (28) we obtain
\begin{equation}
\delta x = - \, { \delta a  \over a^2 } 
\end{equation}
so that, requiring $a(x=0)=g$, we find
\begin{equation}
a(x) =  { g  \over 1+gx } 
\end{equation}
for the position-dependent acceleration that will ensure the rigid motion of the accelerating frame. Substituting into (28) and recalling that $x= X_0$ is a fixed point we have the explicit solution\cite{Moller}
\begin{equation}
X = { 1 \over g }  \left( \sqrt { (1+gx)^2+ g^2 T^2  } -1 \right)
\end{equation}
The motion of the point $x$ is called hyperbolic because (32) implies
\begin{equation}
\left( { 1+gX \over g }  \right)^2 - T^2 =   \left( { 1+gx \over g }  \right)^2 = {1 \over a^2}
\end{equation}
and this represents a hyperbola when drawn in an $X-T$ spacetime diagram.

The relationship between the proper time at $x$ and the inertial time $T$ follows from
\begin{equation}
d \tau (x) = { dT  \over \gamma (x) } = { dT  \over \sqrt {1+ a^2 T^2  } } 
\end{equation}
Thus
\begin{equation}
T = { 1 \over a }  \sinh [ a \, \tau (x)] =  { 1+gx \over g } \,  \sinh \left( {g \, \tau(x) \over 1+gx} \right)
\end{equation}
It is customary (but by no means necessary!) to define a time coordinate for the accelerating frame by
\begin{equation}
t = { \tau (x) \over 1+gx}
\end{equation}
in terms of which (35) simplifies to
\begin{equation}
T = { 1+gx \over g } \,  \sinh g t
\end{equation}
and (32) becomes
\begin{equation}
X = { 1 \over g }  \bigl[ (1+gx) \cosh g t -1 \bigr]
\end{equation}
It is now easy to show that the inertial spacetime interval $ds^2 = dT^2 - dX^2$ takes the Kottler-M\o ller form (10) when written in terms of the coordinates $x$, $t$. Hamilton\cite{Hamilton} has analyzed several interesting phenomena in accelerated frames based on this metric.

A straightforward generalization of the above results starts by noticing that the rigidity condition (29) may be stated in terms of an arbitrary coordinate $x'$. If we set $x = [ \lambda (x') -1 ]/g$, with $ \lambda (0) = 1$ and $d \lambda (x') /dx' \neq 0$, then
\begin{equation}
\delta X(T) =  { 1  \over g }  {  d \lambda  \over dx' } \; { \delta x'  \over \gamma (x') } 
\end{equation}
Note that (39), just like (29), is a relationship between the {\sl proper} length and $\delta X$, so that, despite some claims to the contrary,\cite {Adler} the results below will still represent a rigid frame. Repeating the previous steps leads to
\begin{equation}
a(x') =  { g  \over \lambda (x') } 
\end{equation}
as the appropriate acceleration in this new coordinate $x'$. Similarly, a new time coordinate
\begin{equation}
t' = { \tau (x') \over  \lambda (x')}
\end{equation}
may be defined to turn the solution (37), (38) into\cite{Tilbrook}
\begin{eqnarray}
T \!\!&=&\!\! { \lambda (x') \over g } \,  \sinh g t' \\
X \!\!&=&\!\! { 1 \over g } \, \bigl[ \lambda (x') \cosh g t' -1 \bigr]
\end{eqnarray}
When these expressions are used to write the spacetime interval $ds^2 = dT^2 - dX^2$ in terms of $x'$ and $t'$, one is led back to the form (11) used as a starting point by Rohrlich and Tilbrook, together with the condition (12) derived by both authors from different considerations.

The main lesson of this section is that relativistic effects render the demand of a constant acceleration for all points, despite its intuitive nonrelativistic appeal, inconsistent with our desire to maintain the integrity of the accelerating reference system. The only way to ensure the rigidity of the accelerating frame in special relativity is to have its points move with the acceleration (31) or, more generally, (40). It should now be apparent that, quite apart from the ambiguities inherent in a definition of acceleration in metric theories of gravity, the very notion that ``acceleration $=-g$" is the correct way to specify a uniform gravitational field and its connection to accelerated frames through the equivalence principle is mistaken in the relativistic regime (fact F3).

\section{General relativity and the uniform gravitational field}

In this section we consider the second aspect of the problem by turning our attention to the question of how to implement the concept of a uniform gravitational field in general relativity. It is fairly reasonable to expect that for this type of field it should be possible to find a coordinate system where the metric coefficients are time-independent and the spacetime interval $ds^2$ time-reversal invariant, so that terms such as $dx dt$, $dy dt$, and $dz dt$ do not appear (both properties also follow from the requirement of staticity; see, e.g., Ref.\cite{Rindler}). Furthermore, a uniform field should have translation invariance along planar spatial cross sections of the geometry; we shall take these planes to be orthogonal to the $x$-coordinate, which implies that the metric coefficients can only depend on $x$. In addition to homogeneity we should also require isotropy in each of these planes; the coefficients multiplying $dy^2$ and $dz^2$ must then be identical. We therefore use as our starting point the metric
\begin{equation}
ds^2 = \lambda^2(x) \, dt^2 - \sigma^2(x) \, dx^2- \xi^2(x) \, (dy^2 + dz^2)
\end{equation}
It is worth pointing out that our assumptions can be reduced to just planar homogeneity and isotropy: Taub's theorem\cite{Taub} states that a spacetime with plane symmetry with $R_{\mu \nu} =0$ admits a coordinate system where the line element is static.

The non-vanishing Christoffel symbols for the metric (44) are (primes indicate derivatives with respect to $x$)
\begin{eqnarray}
\Gamma^0_{01} \!\!&=&\!\!  \Gamma^0_{10} = {\lambda' \over \lambda} \\
\Gamma^1_{00} \!\!&=&\!\!  { \lambda \lambda' \over \sigma^2} \\
\Gamma^1_{11} \!\!&=&\!\!  { \sigma' \over \sigma} \\
\Gamma^1_{22} \!\!&=&\!\!  \Gamma^1_{33} = -{ \xi \xi' \over \sigma^2}  \\
\Gamma^2_{12} \!\!&=&\!\!  \Gamma^2_{21} =\Gamma^3_{13} =\Gamma^3_{31} = {\xi' \over  \xi }
\end{eqnarray}
and the resulting nonvanishing components of the Ricci tensor $R_{\mu \nu}$ are
\begin{eqnarray}
R_{00} \!\!&=&\!\! {\lambda \over \sigma^3 \xi} \, \bigl[ 2 \sigma \lambda' \xi' - \xi ( \lambda' \sigma' - \sigma  \lambda'' ) \bigr] \\
R_{11} \!\!&=&\!\! {1\over \lambda \sigma \xi} \, \bigl[ \xi ( \lambda' \sigma' - \sigma  \lambda'' )  + 2 \lambda ( \sigma' \xi' - \sigma  \xi'' ) \bigr] \\
R_{22} \!\!&=&\!\! R_{33} = {1\over \lambda \sigma^3} \, \bigl[ \lambda \xi ( \sigma' \xi' - \sigma  \xi'' ) - \sigma \xi'  ( \xi \lambda' + \lambda  \xi' ) \bigr]
\end{eqnarray}
The Einstein vacuum equations, $R_{\mu \nu} =0$, imply that all the square brackets in (50)-(52) must vanish. Adding the first two brackets gives
\begin{equation}
\sigma \lambda' \xi' +  \lambda ( \sigma' \xi' - \sigma  \xi'' ) =0
\end{equation}
which is equivalent to
\begin{equation}
\lambda^2 \sigma^2 \left(  {\xi' \over \lambda \sigma } \right)' =0
\end{equation}
Hence, with $A$ an integration constant,
\begin{equation}
\xi' = A \lambda \sigma 
\end{equation}
There are now two possibilities. If $A=0$, $\xi$ is constant and we may set $\xi =1$ by appropriately rescaling our $y$ and $z$ coordinates. In this case there is only one independent nontrivial equation left, namely  $\lambda' \sigma' - \sigma  \lambda''=0$. This easily integrated to reproduce equation (12), and we see that, for $A=0$, the metric is of the Rohrlich-Tilbrook form
\begin{equation}
ds^2 = \lambda^2(x) \, dt^2 - \sigma^2(x) \, dx^2- dy^2 - dz^2
\end{equation}
with
\begin{equation}
\lambda' = g \sigma
\end{equation}
As we already know, this represents an accelerated observer in flat spacetime when the integration constant $g \neq 0$, and -- after a suitable redefinition of the $x$ and $t$ coordinates -- a standard inertial observer in flat spacetime when $g=0$.

If $A\neq 0$, we may cast the equation $R_{22}=0$ in terms of the auxiliary variable $\psi = \xi \xi'$ as
\begin{equation}
{\psi' \over \psi } = {\sigma' \over \sigma } - {\lambda' \over \lambda }
\end{equation}
whose solution is
\begin{equation}
\psi= B \, {\sigma \over \lambda }
\end{equation}
Recalling (55) we have
\begin{equation}
\xi = {B \over A \lambda^2 }
\end{equation}
and also
\begin{equation}
\sigma = -{ 2B \over A^2 } \, {\lambda' \over \lambda^4 }
\end{equation}
The reader may verify that no additional independent constraints arise from $R_{\mu \nu} =0$.

We now follow the arbitrary but widely used convention to choose $x=0$ as the plane where the metric takes a Minkowski form, so that $\lambda (0) = \sigma (0) = \xi (0) =1$. This may always be done since it amounts to just a global rescaling by constants of all the coordinates at the plane $x=0$. From (60) we then see that $B=A$ in this convention, and either (55) or (61) may later be used to specify the remaining constant $A$. The resulting metric is
\begin{equation}
ds^2 = \lambda^2 \, dt^2 - \left( {2 \lambda' \over A \lambda^4 }  \right)^2  dx^2 - \lambda^{-4} (dy^2 + dz^2)
\end{equation}
With different choices for the function $\lambda$, this metric has been studied by various authors (see, e.g., Refs.\cite {Amundsen, daSilva} and references therein). Well-known special cases are Taub's solution\cite{Taub} ($\lambda_T = (1 - 2Ax_T)^{-1/4}$) and the Kasner-Das solution\cite{Kasner, Das} ($\lambda_{KD} = (1 - Ax_{KD}/2)^{-1}$). While interesting in its own right, we shall not pursue here a study of this second, $A \neq 0$ alternative any further because the metric (62) represents a geometry that is singular along an infinite two-dimensional spatial cross section. Indeed, a calculation of the Kretschmann scalar $K = R_{\mu \nu \kappa \rho} R^{\mu \nu \kappa \rho}$ shows that
\begin{equation}
K = 12 A^4 \lambda^{12}
\end{equation}
Because $K$ is an invariant, we are free to compute its value in any convenient coordinate system, and it is quite evident from the special cases mentioned above that $K = \infty$ on the plane specified by $x_T =1/2A$ in Taub's coordinate system or $x_{KD} =2/A$ in the Kasner-Das coordinate system. Note that, unlike the Schwarzschild {\sl coordinate} singularity, this is a true singularity of the geometry. Hence, if we add to our present definition of a uniform gravitational field the condition that the geometry should be singularity-free, we find that {\sl in general relativity the only possible representation of a uniform gravitational field is the flat spacetime} (56)-(57).

\section{Does the equivalence principle hold?}

Section III was devoted to a study of accelerated observers in special relativity, i.e., in the absence of any gravitational effects. Section IV developed, in a completely independent fashion, an essentially unique translation of the concept of a uniform gravitational field into the framework of general relativity. An {\sl interpretation} of the idea is, of course, an unavoidable intermediate step since a gravitational field (or force) does not exist in general relativity, but it is hoped that the assumptions of homogeneity and isotropy are minimalistic enough to justify our use of the phrase ``essentially unique."

In this section we investigate whether these results support the expectation arising from the equivalence principle that a uniform gravitational field should be {\sl globally} indistinguishable from an accelerated frame. To this end, consider a freely falling observer in a uniform gravitational field. According to the previous section, the metric is of the form (56)-(57), and the problem can be trivially reduced to two dimensions if we assume that the free fall is initiated with zero velocity components in the $y$ and $z$ directions. The geodesic equation (2) will once again lead to
\begin{equation}
{d t \over d \tau} = {k \over \lambda ^2}
\end{equation}
Hence, recalling (57), (56) becomes
\begin{equation}
{\lambda^4 \over k^2} \, dt^2 = \lambda^2 \, dt^2 - {1 \over g^2} \, d\lambda^2 
\end{equation}
This is a differential equation for $\lambda$,
\begin{equation}
\left( {d \lambda \over dt} \right)^2 = g^2 \lambda^2 \left(1 - {\lambda^2  \over k^2 } \right)
\end{equation}
with solution
\begin{equation}
\lambda (x) = { k  \over \cosh (gt + \theta) }
\end{equation}
where $\theta$ is an integration constant. Given $\lambda$, this equation specifies the trajectory $x(t)$ of the freely falling observer as seen by an observer at rest in the field.

Now, if $X$, $T$ denote the coordinates used by the freely falling observer, we can easily relate length measurements in the two frames at the initial instant if we make some simplifying assumptions. First, we will agree to synchronize the clocks so that $t=0$ implies $T=0$, and the origins of the frames so that $x=0$ agrees with $X=0$ at $t=0$. Second, we will assume that the fall starts from rest, so that $dx/dt =0$ at $t=0$. This means that, {\sl at the initial instant}, $\sigma^2 dx^2 = dX^2$, or, recalling equation (57),
\begin{equation}
\lambda (x(t=0)) = 1 + gX
\end{equation}
where we have chosen the constant of integration so as to conform to our previous convention $\lambda (x =0) =1$. Because $d \lambda /dx$ must be finite and $dx/dt =0$ at $t=0$, it follows that $d \lambda /dt =0$ at $t=0$, which in turn requires $\theta =0$ in (67). Therefore,
\begin{equation}
\lambda (x) = { 1 + gX  \over \cosh gt }
\end{equation}
that is,
\begin{equation}
X = { 1 \over g } \, \bigl[ \lambda (x) \cosh g t -1 \bigr]
\end{equation}

To find the relationship between the times in the two frames, notice that $ds^2 =dT^2$, since $X$ is fixed for the freely falling observer. Thus $d \tau = dT$, and (69) allows us to integrate (64) to find
\begin{equation}
T = { 1 + gX  \over g } \, \tanh gt = { \lambda (x) \over g } \,   \sinh g t
\end{equation}
Equations (70) and (71) prove that the freely falling observer will conclude that an observer at rest in a uniform gravitational field undergoes the exact same hyperbolic motion as did the observer with constant proper acceleration in flat spacetime when viewed by the inertial observer of Section III. This equivalence is exact: the trajectories (42)-(43) and (70)-(71) are identical for all points and all times where the coordinates $x$ and $t$ are valid (the coordinates in (37)-(38), for instance, will fail for $x < -1/g$, but this is irrelevant here because the same failure will occur in the gravitational case, and because the extension of the coordinate patch is finite rather than infinitesimal). And while it is true that the resulting acceleration does not take the same constant value everywhere in space -- indeed, here $a(x) = g / \lambda (x)$, just as in Eq. (40) -- this should be a welcome consequence since we now know from Section III that this Newtonian preconception about accelerated frames must be abandoned in the relativistic regime (fact F3). Moreover, the non-constant value of the acceleration in this particular situation is emphatically {\sl not} the reason why the equivalence principle applies, in general, only at the (infinitesimal) local level.\cite{Adler} After all, (70) and (71) transform the metric (56) into a Minkowski metric over a {\sl finite} domain. For a detailed exposition of the inextricable link between equivalence and locality in the general case, see, e.g., Sections 9.6 and 9.9 in M\o ller\cite{Moller}.

In summary, much of the confusion in the literature arises from ignoring some of the most basic lessons of general relativity such as the role of coordinates as mere labels, and from trying to impose the nonrelativistic ``acceleration $=-g$" definition on a uniform gravitational field in the relativistic regime. The result is an apparent ``failure" of the equivalence principle. However, as we have shown in this paper, once a few facts about relativistic physics are properly taken into account, it is easy to see that in a uniform gravitational field the spacetime is, as expected, flat, and that the equivalence principle, when correctly understood, is alive and well.



\begin{thebibliography}{99}

\bibitem{Moller}  C. M\o ller, \textit{The Theory of Relativity}, 2nd. ed. (Clarendon Press, Oxford, 1972).

\bibitem{Schutz}  B. F. Schutz, \textit{A First Course in General Relativity}, (Cambridge University Press, Cambridge, 1992).

\bibitem{MTW}  C. W. Misner, K. S. Thorne, and J. A. Wheeler, \textit{Gravitation}, 2nd. ed. (W.H. Freeman, N.Y., 1973).

\bibitem{Stephani}  H. Stephani, \textit{Relativity: An Introduction to Special and General Relativity}, (Cambridge University Press, Cambridge, 2004).

\bibitem{Rindler}  W. Rindler, \textit{Relativity: Special, General, and Cosmological}, 2nd. ed. (Oxford University Press, 2001).

\bibitem{Carroll}  S. Carroll, \textit{Spacetime and Geometry: An Introduction to General Relativity}. (Pearson Addison Wesley, San Francisco, 2004).

\bibitem{Synge}  J. L. Synge, \textit{Relativity: The General Theory}. (North-Holland Publishing Co., Amsterdam, 1960).

\bibitem{Fock}  V. Fock, \textit{Theory of Space, Time, and Gravitation}. (Pergamon Press, London, 1959).

\bibitem{Ohanian}  H. C. Ohanian and R. Ruffini, \textit{Gravitation and Spacetime}, 2nd. ed. (W. W. Norton \& Company, New York, 1994).

\bibitem{Pauli}  W. Pauli, \textit{Theory of Relativity}. (Pergamon Press, London, 1958).

\bibitem{FultonRohrlich}  T. Fulton and F. Rohrlich, ``Classical Radiation from a Uniformly Accelerated Charge'', Ann. Phys. (NY), \textbf{9}, 499-517 (1960).

\bibitem{Narozhny}  N. B. Narozhny, A. M. Fedotov, B. M. Karnakov, V. D. Mur, and V. A. Belinskii, ``Boundary conditions in the Unruh problem'', Phys. Rev. D \textbf{65}, 025004-1--025004-23 (2002).

\bibitem{FullingUnruh} S. A. Fulling and W. G. Unruh, ``Comment on ``Boundary conditions in the Unruh problem''", Phys. Rev. D \textbf{70}, 048701-1--048701-4 (2004).

\bibitem{Crispino}  L. C. B. Crispino, A. Higuchi, and G. E. A. Matsas, ``The Unruh effect and its applications'', Rev. Mod. Phys. \textbf{80}, 787-838 (2008).

\bibitem{Rohrlich}  F. Rohrlich, ``The Principle of Equivalence'', Ann. Phys. (NY) \textbf{22}, 169-191 (1963).

\bibitem{Desloge}  E. A. Desloge, ``Nonequivalence of a uniformly
accelerating reference frame and a frame at rest in a uniform gravitational
field'', Am. J. Phys. \textbf{57}, 1121-1125 (1989).

\bibitem{Gron} Gr{\o }n and E. Eriksen, ``Equivalence in Two-,
Three-, and Four-dimensional Space-Times'', Int. J. Theo. Phys. \textbf{31}, 1421-1432 (1992).

\bibitem{Tilbrook}  D. Tilbrook, ``General coordinatisations of the flat
space-time of constant proper-acceleration'', Aust. J. Phys. \textbf{50},
851-868 (1997).

\bibitem{Kottler1}  F. Kottler, ``Relativit\"atsprinzip und beschleunigte Bewegung'', Annalen der Physik \textbf{44}, 701-748 (1914).

\bibitem{Kottler2}  F. Kottler, ``\"Uber die physikalischen Grundlagen der Einsteinschen Gravitationstheorie'', Annalen der Physik \textbf{56}, 401-462 (1918).

\bibitem{Whittaker}  E. T. Whittaker, ``On electric phenomena in a gravitational field'', Proc. Roy. Soc. A \textbf{116}, 720-735 (1927).

\bibitem{Adler} C. G. Adler and R. W. Brehme, ``Relativistic solutions to
the falling body in a uniform gravitation field", Am. J. Phys. {\bf 59}, 209-213(1991).

\bibitem{Hamilton} J. D. Hamilton, ``The uniformly accelerated reference frame", Am. J. Phys. {\bf 46}, 83-89 (1978).

\bibitem{Amundsen}  P. A. Amundsen and {\O }. Gr{\o }n, ``General static
plane-symmetric solutions of the Einstein-Maxwell equations'', Phys. Rev. D 
\textbf{27}, 1731-1739 (1983).

\bibitem{daSilva}  M. F. A. da Silva, A. Wang, and N. O. Santos, ``On the sources of static plane vacuum spacetimes'', Phys. Lett. A \textbf{244}, 462-466 (1998).

\bibitem{Taub} A. H. Taub, ``Empty Space-Times Admitting a Three Parameter Group of Motions", Phys. Rev. {\bf 53}, 472-490 (1951).

\bibitem{Kasner} E. Kasner, ``Solutions of the Einstein Equations Involving Functions of Only One Variable", Trans. Amer. Math. Soc. {\bf 27}, 155-162 (1925).

\bibitem{Das} A. Das, ``Static Gravitational Fields. I. Eight Theorems", J. Math. Phys. {\bf 12}, 1136-1142 (1971).



\end{thebibliography}
\end{document}